\documentclass[a4paper,11pt]{article}
\usepackage{pos}
\usepackage{physics}
\usepackage{slashed}
\usepackage{bm}
\usepackage{subcaption}
\usepackage{dirtytalk}
\usepackage{tikz}
\usepackage[compat=1.0.0]{tikz-feynman}

\title{
\vspace*{-0.75cm}
\begin{minipage}{\textwidth}
\begin{flushright}
\texttt{\footnotesize
PoS(LATTICE2026)073\\
DESY-26-159\\
}
\end{flushright}
\end{minipage}\\[15pt]
\vspace*{+0.75cm}
Optimised Distillation for Nucleon Correlation Functions}
\ShortTitle{Optimised Distillation for Nucleon Correlation Functions}

\author*[a]{J.~A.~Crawford}
\author[a]{J.~Finkenrath}
\author[b,c]{L.~Barca}

\affiliation[a]{
Department of Physics, Bergische Universität Wuppertal, Gaußstr. 20, 42119 Wuppertal, Germany
}
\affiliation[b]{Aix Marseille Univ, Universit\'e de Toulon, CNRS, CPT, Marseille, France}
\affiliation[c]{John von Neumann-Institut für Computing NIC, Deutsches Elektronen-Synchrotron DESY, Germany}

\emailAdd{crawford@uni-wuppertal.de}

\abstract{The calculation and analysis of nucleon correlation functions is essential for understanding the observed spectra and structure of these ubiquitous hadrons. Two challenges arise in their analysis: at early Euclidean times, they are dominated by excited state contributions and at later Euclidean times, the signal-to-noise ratio degrades exponentially. This leads to a common challenge in identifying a reasonable window on which to extract the ground state energy. By modulating the contribution of each Laplacian eigenmode to the correlator via distillation profiles, we build optimised operators which improve the overlap with the nucleon ground state. In these proceedings, we present our progress on constructing optimised distillation profiles for nucleons and their excited state suppression in two- and three-point correlation functions.}

\FullConference{The 43rd International Symposium on Lattice Field Theory (Lattice 2026)\\
July 26 to August 1, 2026\\
University of Maryland, College Park, USA\\}

\begin{document}
\maketitle
\section{Introduction}
The calculation of Euclidean two-point correlation functions in lattice QCD provides access to the hadronic energy spectrum and three-point correlation functions provide access to hadronic matrix elements. However, in most channels, the analysis of these correlation functions is complicated by a windowing problem. At early Euclidean times, the correlator signal is dominated by excited states, whereas at late Euclidean times, the ground state signal is lost to noise. The determination of a region in which the ground state dominates the signal at an acceptable level of statistical precision is therefore a non-trivial task.
\par

Distillation \cite{HadronSpectrum:2009krc} is a powerful framework in which to compute hadronic correlation functions. Working within the subspace spanned by the low modes of the 3D gauge-covariant Laplacian,
\begin{equation}
    \nabla^2_{xy}[t] = 6\delta_{xy}-\sum_{j=1}^3\left( U_j(x,t)\delta_{x+\hat{j},y} + U^\dagger(x-\hat{j},t)\delta_{x-\hat{j},y}\right)
\end{equation}
where the eigenvectors on each timeslice $t$ are
\begin{equation}
    -\nabla^2_{xy}[t] v^j[\bm{y},t] = \lambda^j[t]v^j[\bm{x},t].
\end{equation}
Higher eigenmodes of the 3D gauge-covariant Laplacian are exponentially suppressed in conventional smearing algorithms, and so implementing a cut-off in the number of eigenmodes retained in the subspace, $j\leq N_v$, defines an effective smoothing algorithm. Outside of volume-scaling arguments, it is not clear beforehand how many eigenvectors should be kept to achieve the optimal overlap with the state of interest. The use of distillation profiles \cite{Urrea-Nino:2022bji, Urrea-Nino:2025afu, Urrea-Nino:2026cjj} negates this issue by reweighting each eigenmode such that it contributes optimally to the state of interest. Calculations of correlation functions within this subspace are exact and hence statistical variation is reduced compared to standard correlation function calculations.
\par

Variational methods are often used in lattice calculations to enhance overlaps with states of interest \cite{Blossier:2009kd}. In the context of nucleon matrix element calculations, excited state contamination is arguably the leading systematic uncertainty. The use of a variational smearing basis has previously been explored \cite{Yoon:2016dij, Dragos:2016rtx}, showing suppression of excited state contamination across several channels. However in certain channels, multi-particle states can dominate the excited state contamination and dedicated multi-particle operators are needed to suppress these contaminations \cite{Barca:2025det}. In this work, we present our progress on a cost-effective means for generating a variational basis within the distillation framework on which to analyse nucleon two- and three-point correlation functions.

\section{Methods}
\subsection{Distillation Profiles}
Using the distillation framework, we construct nucleon correlation functions from re-usable components. Perambulators, $\tau$, replace propagators by projecting the propagator into the space spanned by the $N_v$ lowest modes of the 3D gauge-covariant Laplacian,
\begin{equation}
    \tau^{ij}_{\alpha\beta}(t_2,t_1) = v^{\dagger\,i}_a[t_2]D^{-1}(t_2,t_1)^{ab}_{\alpha\beta}v^{j}_{b}[t_1],
\end{equation}
where $v^i$ are the Laplacian eigenvectors. Elementals, $\Phi$, replace interpolating operators that capture the colour and Dirac structure of the nucleon at rest,
\begin{equation}
    \Phi^{ijk}_{\alpha\beta}[t] = \sum_{\bm{x}}\epsilon^{abc}v^{i}_a[\bm{x},t]v^{j}_b[\bm{x},t]v^{k}_c[\bm{x},t] [C\gamma_5]_{\alpha\beta},
\end{equation}
where $\epsilon^{abc}$ is 3D Levi-Civita symbol, with $\epsilon^{123}=+1$ and $C=i\gamma_2\gamma_4$ is the charge-conjugation matrix. In the implementation of distillation profiles, we further modulate the relative weight of each eigenmode via a Gaussian profile,
\begin{equation}
    v^i[t] \to \exp\left( -\frac{\lambda_i[t]^2}{2\sigma^2}\right)v^i[t],
\end{equation}
where $\sigma$ controls the width of the Gaussian profile. Contrasted with standard distillation, where the weight of each eigenmode is equal, the Gaussian profiles place a higher weight on lower eigenmodes and suppress higher eigenmodes. By using multiple widths, we effectively construct a variational smearing basis as shown in Refs. \cite{Urrea-Nino:2022bji, Urrea-Nino:2025afu, Urrea-Nino:2026cjj}. Importantly, this does not require any additional inversions and is performed entirely at the contraction level. We solve a generalised eigenvalue problem (GEVP) on each timeslice $t$ relative to a reference timeslice $t_0$,
\begin{equation}
    C(t)v^{(u)}(t,t_0) = \lambda^{(u)}(t,t_0)C(t_0)v^{(u)}(t,t_0),
\end{equation}
where the eigenvalues $\lambda^{(u)}(t,t_0) = e^{-E_u(t-t_0)}$, with $E_u$ being the energy of state $u$, define the principle correlators and the eigenvectors $v^{(u)}(t,t_0)$ define the optimal linear combinations of the variational basis of distillation profiles. Once the eigenvectors are calculated, the correlation matrix is transformed to the eigenbasis, with the diagonal entries corresponding to the tower of energy eigenstates.

\subsection{Three-Point Function Construction}
\begin{figure}[t!]
    \centering
    \begin{tikzpicture}
        \begin{feynman}
            \vertex (a) at (0,0);
            \node[crossed dot] (b) at (5,1);
            \vertex (d) at (5,2) {\(\Gamma_{\alpha\beta}\)};
            \vertex (c) at (10,0);
            \diagram*{  
            (a) -- [fermion, out=30, in=180] (b) -- [fermion, out=0, in=150] (c),
            (d) -- [photon] (b),
            };
        \end{feynman}
        \draw[fill=black] (0,0) circle (2pt);
        \draw[fill=black] (10,0) circle (2pt);
        \node[anchor=north] at (0,0) {\(t_1\)};
        \node[anchor=north] at (10,0) {\(t_2\)};
        \node[anchor=north] at (5,0.9) {\(\tau\)};
        \node[anchor=south east] at (0,0) {\(v^{i\dagger}[t_1]\)};
        \node[anchor=south west] at (10,0) {\(v^{j}[t_2]\)};
        \node[anchor=south east] at (5,1) {\(\eta^{k'}[\tau]\)};
        \node[anchor=south west] at (5,1) {\(\eta^{k'\dagger}[\tau]\)};
    \end{tikzpicture}
    \caption{Stochastic estimation of a local current insertion.}
    \label{fig: stochambulator}
\end{figure}
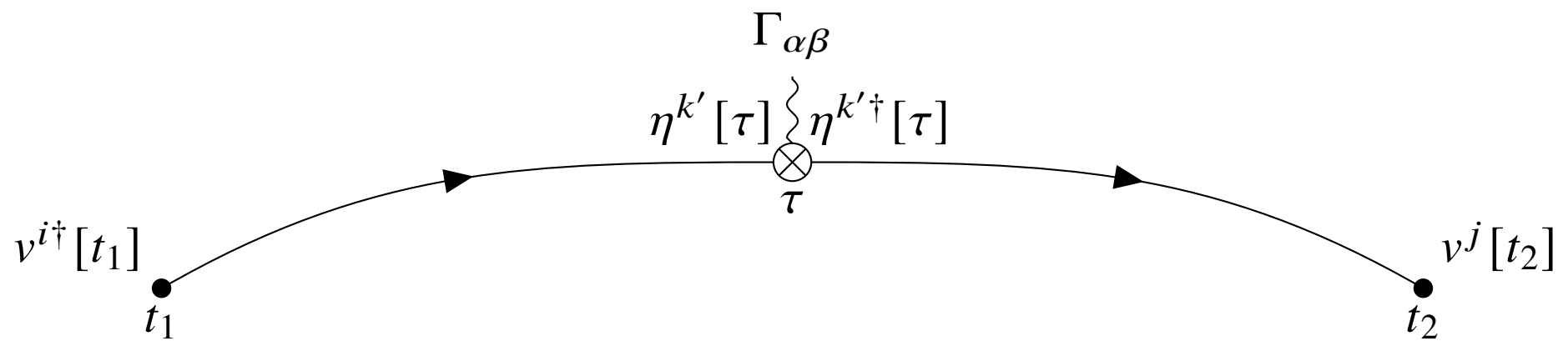

We use a stochastic method to calculate the generalised perambulators used to construct three-point functions. We first define \say{stochambulators}, where rather than projecting onto a Laplacian eigenvector, we instead project the propagators onto a $Z_4$ stochastic vector $\eta^{j'}$ at one end,
\begin{equation}
    \tau^{ij'}_{\alpha\beta}(t_1,t_2)=v^{i\dagger}_a[t_1]D^{-1}(t_1,t_2)^{ab}_{\alpha\beta}\eta^{j'}_b[t_2],
\end{equation}
where we use primed indices to indicate stochastic indices and unprimed indices to indicate distillation indices. The generalised perambulator for an insertion $\Gamma$ is estimated stochastically as
\begin{equation}
    G^{ij}_{\alpha\beta}(\Gamma;t_1,\tau,t_2)=\frac{1}{N_s}\sum_{k'=1}^{N_s}\tau^{ik'}_{\alpha\alpha'}(t_1,\tau)\Gamma_{\alpha'\beta'}\tau^{k'j}_{\beta'\beta}(\tau,t_2).
\end{equation}
This contraction is shown schematically in Figure \ref{fig: stochambulator}. In the limit $N_s\to\infty$, the stochastic estimator of the current insertion converges to the local current insertion, and its variance scales as $1/N_s$.
Extension of our method to also incorporate blended exact/stochastic estimators of generalised perambulators \cite{Hu:2025vhd, Wang:2025nsd} is left to future work.

\section{Results}
We perform our analysis on a single CLS lattice ensemble, A653, which simulates $N_f=3$ flavours of non-perturbatively $\mathcal{O}(a)$ improved Wilson fermions \cite{SHEIKHOLESLAMI1985572, Bulava:2013cta} and a tree-level Symanzik improved gauge action \cite{WEISZ19831}. The ensemble lies at the SU(3)-flavour-symmetric point, with degenerate $u$, $d$ and $s$ quark masses set to $m_{\rm{symm}}=(2m_l+m_s)/3$, yielding a pion mass of approximately $m_\pi \approx 429$ MeV. Details of the lattice ensemble are shown in Table \ref{tab:lattices}.
\begin{table}[t!]
            \centering
            \caption{Details of the CLS lattice used in this analysis. $L$ ($T$) is the spatial (temporal) extent of the lattice, $\beta$ is the inverse gauge coupling, $a$ is the lattice spacing, $m_\pi$ is the pion mass, $N_{\rm{cfg}}$ is the number of configurations used in our analysis, $N_v$ is the number of eigenvectors used to construct the distillation subspace and $N_s$ is the number of stochastic vectors used at the local current insertion.}
            \label{tab:lattices}
            \begin{tabular}{c|c c c c c c c}
                id & $L^3 \times T$ & $\beta$ & $a$ (fm) & $m_\pi$ (MeV) & $N_{\text{cfg}}$ & $N_v$ & $N_s$\\
                \hline\hline
                 A653 & $24^3\times 48$ & 3.34 & 0.098 & 429 & 132 & 50 & 50\\
                 \hline
            \end{tabular}
\end{table}

\subsection{Nucleon Two-Point Functions} 
\begin{figure}[t!]
\centering
\begin{subfigure}[t]{0.49\textwidth}    
    \centering
    \includegraphics[width=0.95\linewidth]{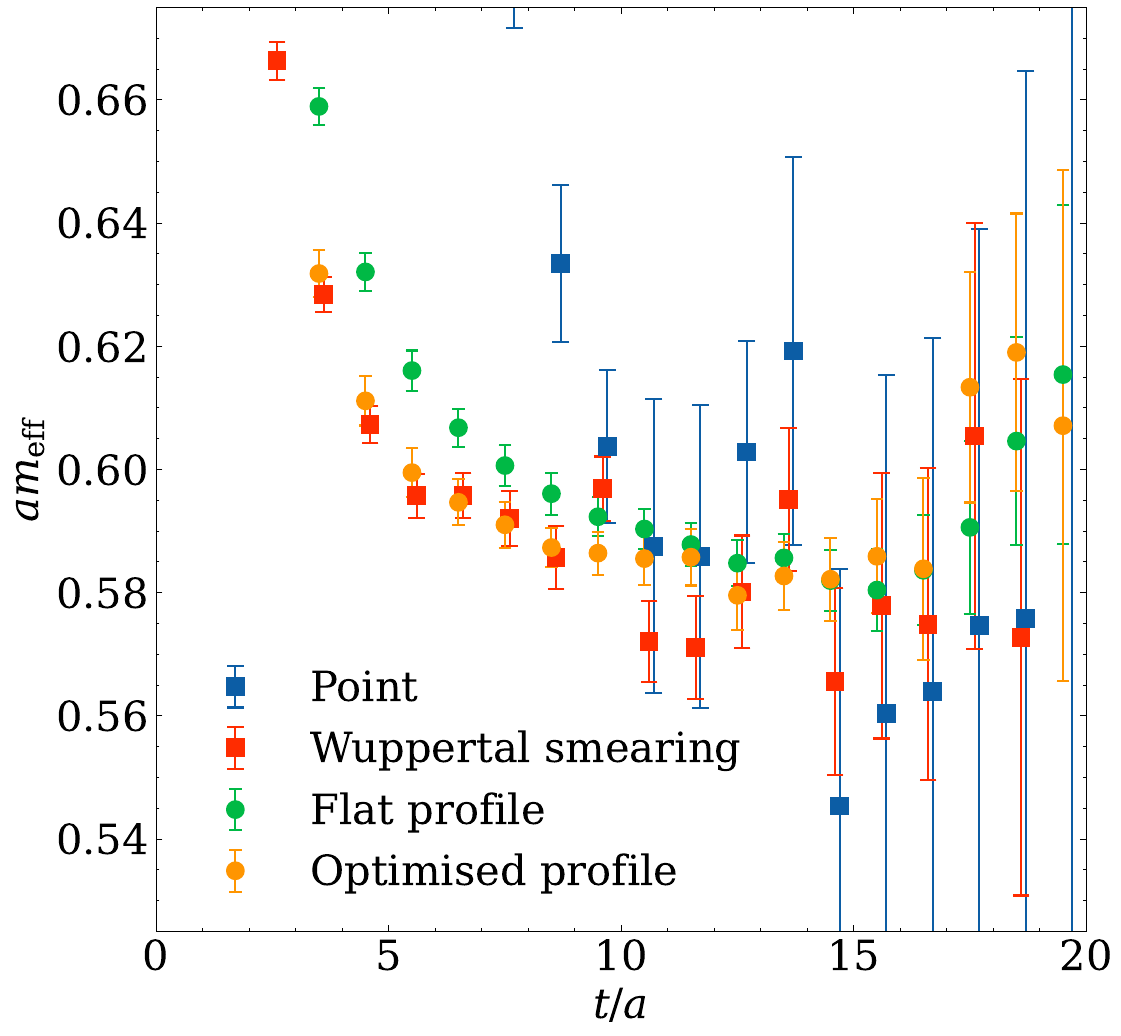}
    \caption{Comparison of different smearing methods for the nucleon effective mass.}
    \label{fig: nucleon effective mass}
\end{subfigure}
\hfill
\begin{subfigure}[t]{0.49\textwidth}
    \includegraphics[width=0.89\linewidth]{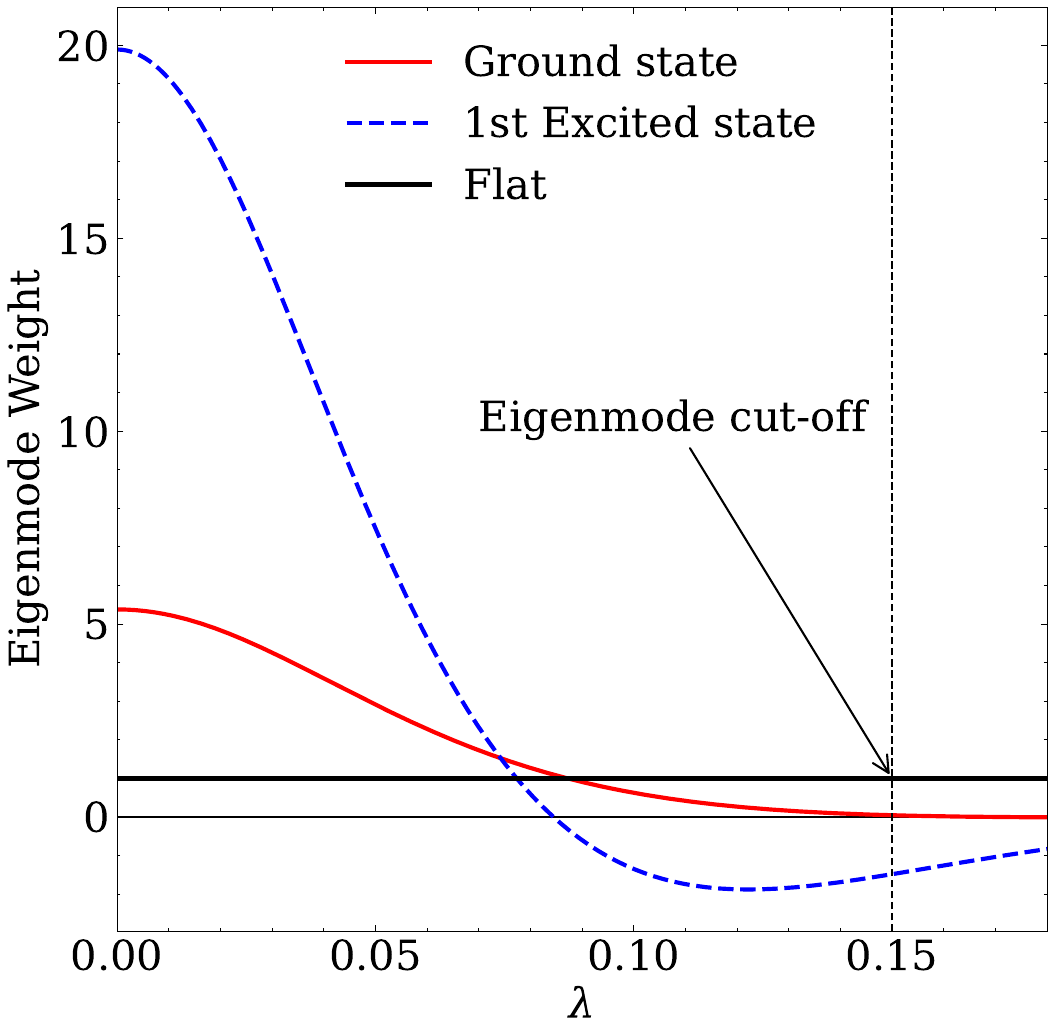}
    \caption{Optimised profiles for the nucleon at rest.}
    \label{fig: nucleon optimised profiles}
\end{subfigure}
\caption{Effective masses and distillation profiles for a nucleon at rest.}
\label{fig: nucleon at rest}
\end{figure}

We compare the performance of nucleon correlators constructed using distillation against conventional correlators in Figure \ref{fig: nucleon at rest}. We first highlight the excited state behaviour of point-to-point correlators (blue squares). Whilst using a factor of 12 less statistics, these results qualitatively show the excited state suppression achieved from optimal smearing. The red squares represent nucleon correlators constructed from conventional propagators between smeared sources and sinks. These smeared correlators are constructed on an equal-cost basis as our distillation results. The smearing parameters of the Wuppertal smearing \cite{Gusken1990} at the source and sink, as well as the APE gauge link smearing \cite{APE:1987ehd}, were tuned to optimise the overlap with the nucleon ground state, see Ref.~\cite{RQCD:2022xux} for the specific smearing parameters. The optimised smearing yields ground state dominance much earlier in Euclidean time compared to the point-to-point correlators.
\par

When constructing correlation functions in the distillation subspace, the number of retained eigenvectors $N_v$ acts as an effective smearing parameter. The distilled correlators constructed with flat profiles (green circles) show a slower growth in statistical noise when compared to the optimised Wuppertal smearing parameters. Evidently, for our choice of $N_v=50$ eigenvectors, the spatial smearing of the nucleon was not optimal, as some excited state contamination is present. By applying our basis of Gaussian profiles to the set of $N_v$ eigenvectors, we perform a GEVP to optimise the weighting of each of the retained eigenmodes. The optimal weighting of each eigenmode for constructing the ground state nucleon is shown in Figure \ref{fig: nucleon optimised profiles}. Using the eigenvectors from the GEVP, the optimised nucleon correlator is constructed from
\begin{equation}
    C(t-t_0) = v^{(0)\dagger}_i(t,t_0)C_{ij}(t)v^{(0)}_j(t,t_0),
\end{equation}
where $v^{(0)}(t,t_0)$ is the eigenvector corresponding to the lowest eigenvalue. Compared to the flat profile, the optimal weighting places higher weight to lower eigenmodes, decreasing to zero as the modes approach the cut-off. The corresponding correlator constructed from this optimal weighting of eigenmodes is shown as the yellow circles in Figure \ref{fig: nucleon effective mass}. The excited state suppression is comparable to the optimised Wuppertal smearing, indicating that the spatial smearing from this combination of eigenmodes is comparable to the physical nucleon wavefunction. However, as we are working in the distillation subspace, the statistical noise of the correlator is more controlled.
\par

\begin{figure}[t!]
\centering
\begin{subfigure}[t]{0.49\textwidth}    
    \centering
    \includegraphics[width=0.90\linewidth]{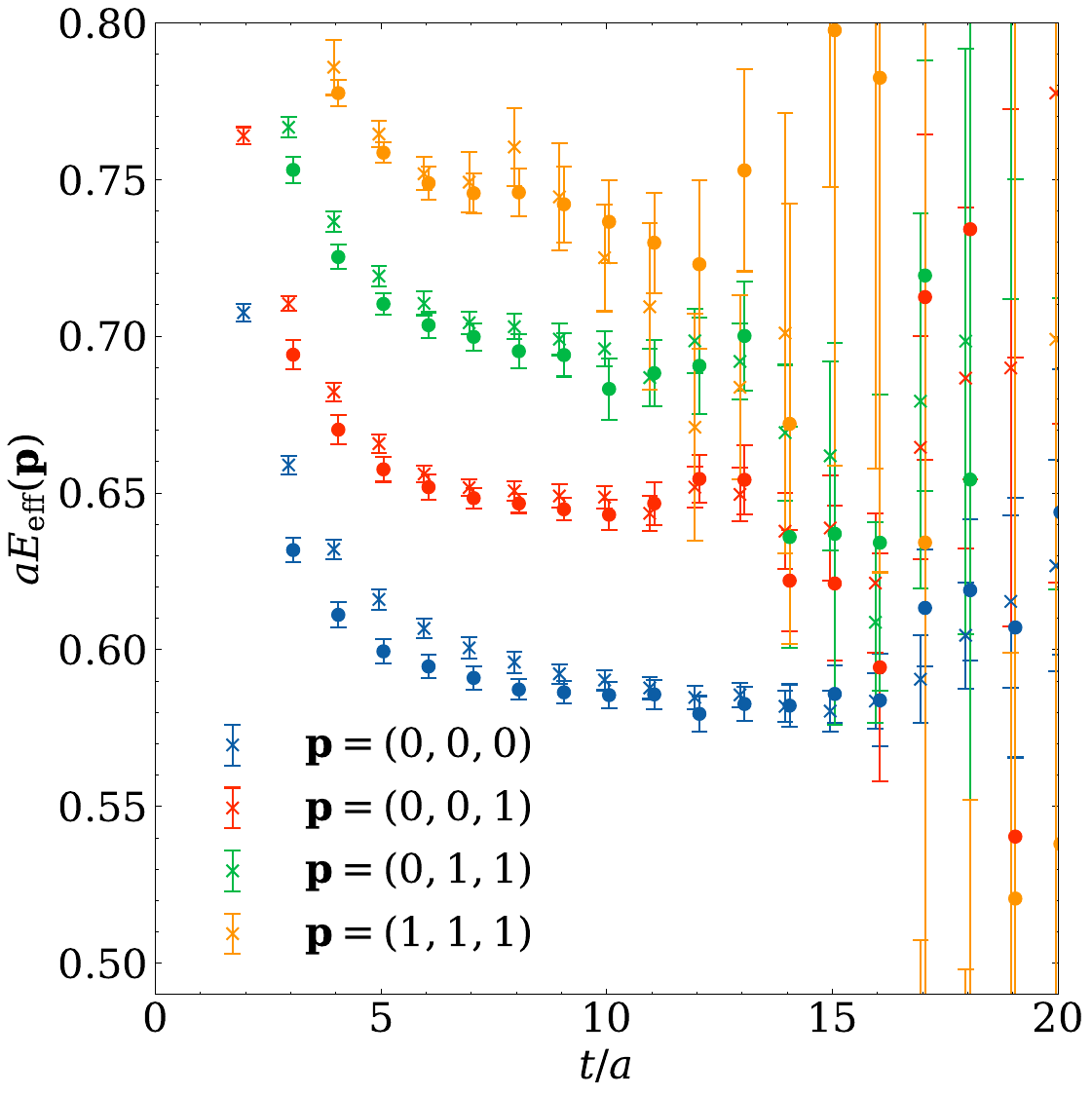}
    \caption{Comparison of different smearing methods for the nucleon effective mass.}
    \label{fig: boosted eff masses}
\end{subfigure}
\hfill
\begin{subfigure}[t]{0.49\textwidth}
    \includegraphics[width=0.88\linewidth]{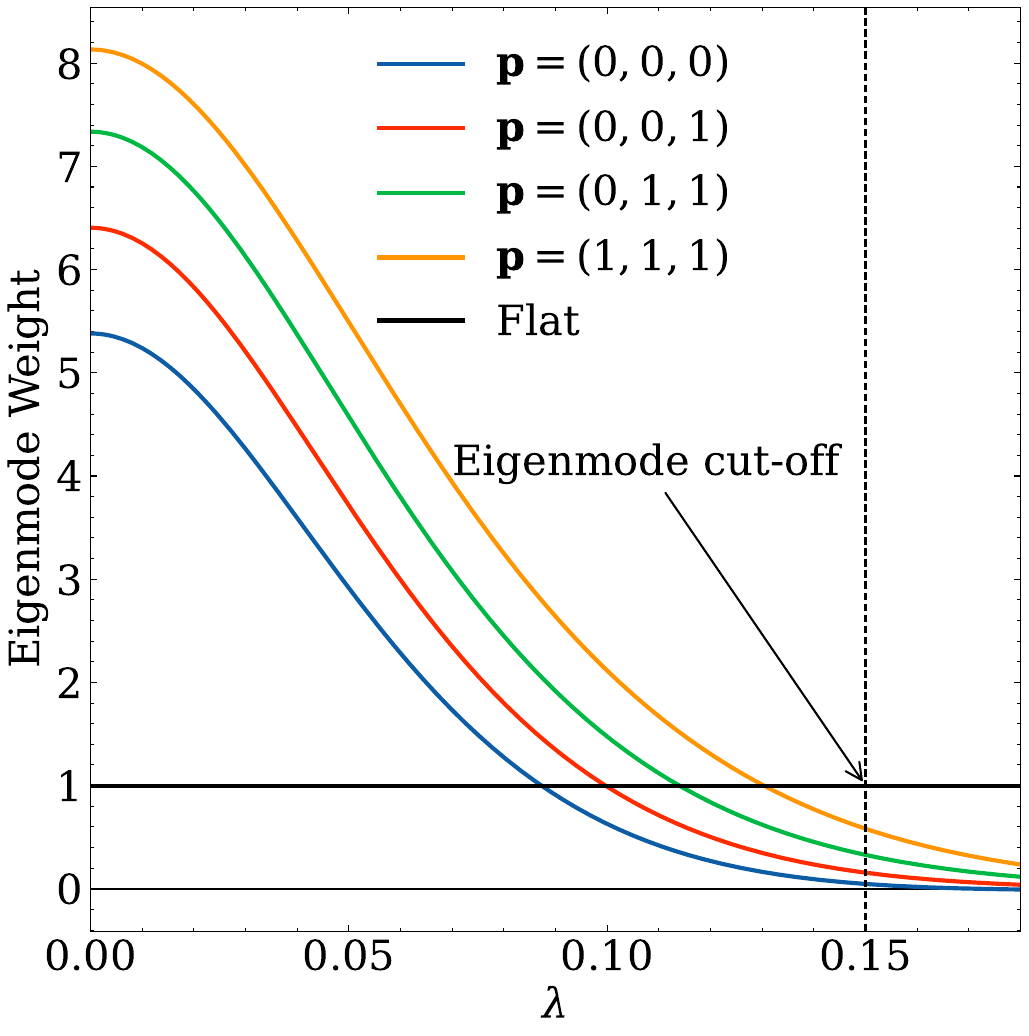}
    \caption{Optimised profiles for boosted nucleons.}
    \label{fig: boosted profiles}
\end{subfigure}
\caption{Effective masses and distillation profiles for a boosted nucleon.}
\label{fig: boosted nucleon}
\end{figure}

In Figure \ref{fig: boosted nucleon}, we extend our analysis to nucleons with finite momentum. In Figure \ref{fig: boosted eff masses}, we show a comparison between nucleon correlators constructed both using flat profiles (crosses) and with optimised profiles (circles) for three different momentum, as well as the rest case for reference. Momentum is introduced via a Fourier phase in the elemental,
\begin{equation}
    \Phi^{ijk}_{\alpha\beta}(\bm{p},t) = \sum_{\bm{x}}e^{i\bm{p}\cdot\bm{x}}\epsilon^{abc}v_a^i[\bm{x},t]v_b^j[\bm{x},t]v_c^k[\bm{x},t][C\gamma_5]_{\alpha\beta}.
\end{equation}
As the momentum increases, the onset of statistical noise occurs earlier in Euclidean time. We find that the degree of excited state suppression decreases with increasing momentum. From Figure \ref{fig: boosted profiles}, we see that as the nucleon momentum increases, higher eigenmodes receive more weight, including those which go beyond our cut-off value. There are alternative methods for improving the signal quality of finite momentum nucleon correlators, including the use of momentum smearing and phased distillation eigenvectors \cite{Bali:2016lva, Egerer:2020hnc}. The implementation of distillation profiles is perfectly compatible with these techniques. Similar improvements in the nucleon sector were reported by J.~Urrea-Ni\~no \textit{et. al.} at this conference, as well as improvements in the charmonium and glueball sectors \cite{Urrea-Nino:2026nkq}. 

\subsection{Nucleon Three-Point Functions}
By taking a ratio of three-point correlation functions to two-point correlation functions, one can access nucleon matrix elements of the inserted current,
\begin{equation}
\label{eq: flat profile ratio}
    \mathcal{R}_\Gamma(\tau) = \frac{C_{\rm{3pt}}(\tau,t;\Gamma)}{C_{\rm{2pt}}(t)} \xrightarrow[t-\tau \gg 0]{\tau \gg 0} g_\Gamma + \frac{A_1}{A_0}(g_\Gamma^{1\to0} e^{-\Delta E \tau} + g_{\Gamma}^{0\to1}e^{-\Delta E(t-\tau)}) + \ldots
\end{equation}
where $A_{0,1}$ contain the relevant overlap factors, $\Delta E$ is the energy separation between the two states, $g_\Gamma$ is the forward nucleon matrix element $g_\Gamma = \langle p,s | \overline{\psi}\Gamma\psi | p,s\rangle$, and $g_\Gamma^{0\to1}$ and $g_\Gamma^{1\to0}$ indicate the corresponding transition matrix elements between the ground and first excited state. Further suppressed contributions are indicated by the ellipses. By increasing the overlap with the ground state, $A_0$, and decreasing the overlap with the first excited state, $A_1$, we effectively suppress these contributions from these transition matrix elements.
\par

\begin{figure}[t!]
    \centering
    \includegraphics[width=0.85\linewidth]{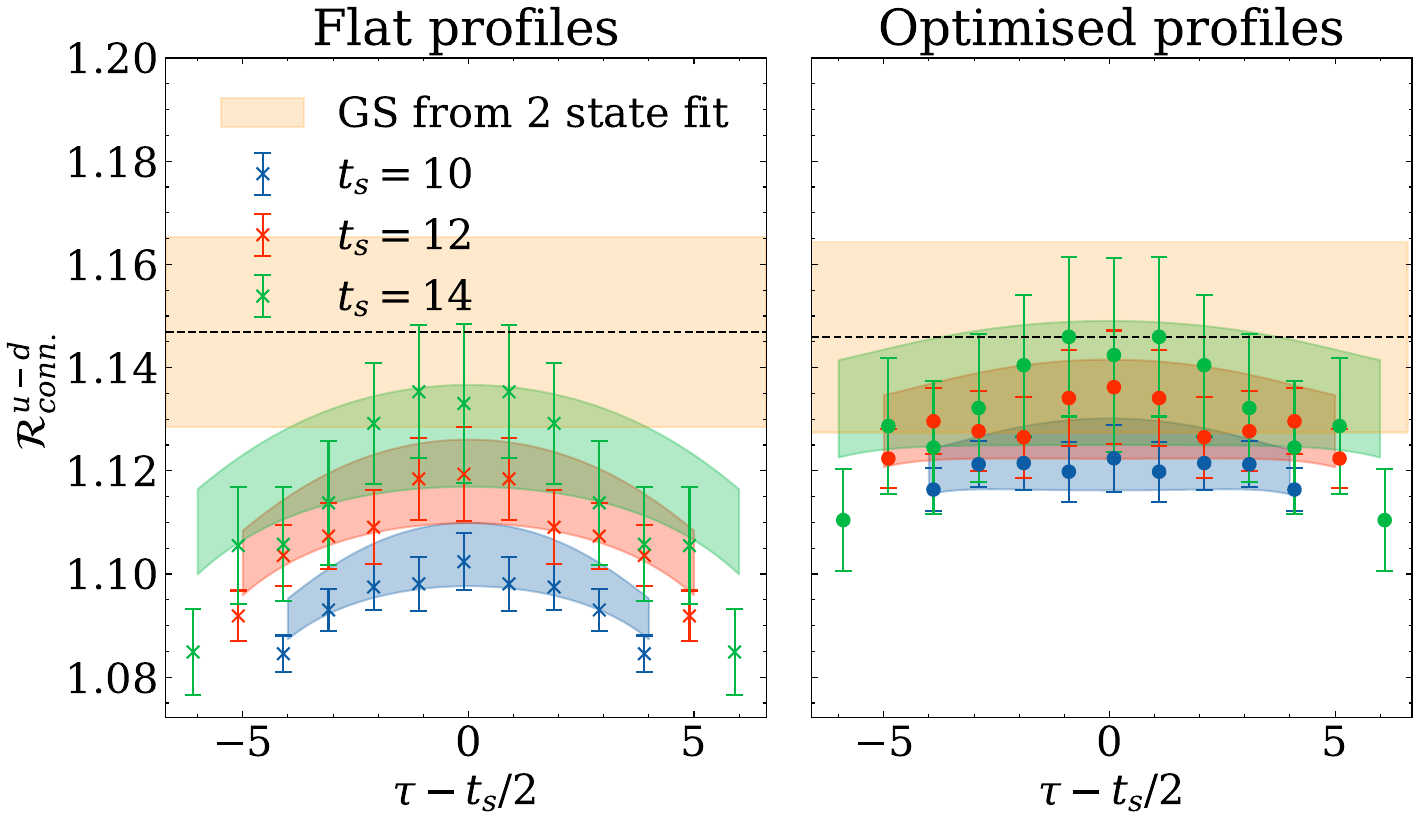}
    \caption{Comparison of $g_A^{u-d}$ extraction using conventional distillation (flat profiles) and with optimised profiles.}
    \label{fig: 3pt comparison}
\end{figure}

Using the same basis of distillation profiles, we construct three-point correlation functions with an axial current insertion. We take the eigenvectors calculated from the two-point correlation function and use these to rotate the three-point function to the same basis. In Figure \ref{fig: 3pt comparison}, we show a comparison of the calculation of $g_A^{u-d}$ using flat profiles, i.e. Equation \eqref{eq: flat profile ratio}, and the same ratio with optimised profiles. On the left plot, the ratio constructed from flat profiles shows a degree of excited state contamination, evidenced by the curvature in $\tau$ and dependence on $t_s$. To extract the matrix element $g_\Gamma$, we first independently fit the two-point function with the ansatz
\begin{equation}
    C_{\rm{2pt}}(t) = |Z_0|^2 e^{-E_0 t}\bigg(1+ \frac{|Z_1|^2}{|Z_0|^2} e^{-\Delta E t}\bigg),
\end{equation}
where $Z_{0,1}$ are the overlap factors and $\Delta E = E_1-E_0$. We similarly use a two-state ansatz for the three-point function,
\begin{equation}
    C_{\rm{3pt}}(\tau,t;\Gamma) = B_{00}e^{-E_0t} + B_{01}(e^{-\Delta E(t-\tau)} + e^{-\Delta E \tau}) + B_{11}e^{-E_1 t},
\end{equation}
where $B_{ij} = Z_i Z_j^*\langle i|\overline{\psi}\Gamma\psi|j\rangle$ with $i,j\in \{0,1\}$. We obtain the fit parameters $B_{ij}$ by fitting directly to the ratio, using the pre-fit parameters from the two-point functions. The resulting fits are shown as the coloured bands. For the flat profiles, we obtain a value of $g_A^{u-d}=1.147(18)$. Contrast this with the right plot using the optimised profiles, where the $\tau$ and $t_s$ dependence are heavily suppressed. We similarly perform a two-state fit to the optimised profile ratio values and obtain $g_A^{u-d}=1.146(18)$, which is consistent with the flat profile case. There is still some residual excited state contamination, however it is more heavily suppressed. Note that one could equivalently perform a plateau fit to largest source-sink separation and obtain a value statistically consistent with the full two-state fit, but with reduced variance.

\section{Summary and Conclusion}
In these proceedings, we have presented an inexpensive method for optimising nucleon operators in the calculation of nucleon correlation functions using distillation. We investigated the efficacy of excited state suppression through a variational approach in the space of these distillation profiles. For a nucleon at rest, we found that the use of optimised distillation profiles recreated the excited state behaviour seen in conventional correlator calculations where the source smearing was tuned for optimal overlap with the nucleon ground state. At finite momentum, the suppression of excited states achieved through optimised profiles decreased with increasing $\bm{p}^2$. In the calculation of nucleon matrix elements, the eigenvectors obtained from the GEVP of the two-point correlator were used to rotate the three-point function to the optimised basis. In the case of the nucleon isovector axial charge $g_A^{u-d}$, we found a considerable reduction in excited state contamination by using these optimised operators. The implementation of distillation profiles is compatible with existing correlator construction methods, and can be combined with other operator bases to construct even larger variational bases. In future work, we plan to investigate other forward nucleon matrix elements and extend our analysis to finite momentum transfer, reducing excited state contamination for form factor calculations.

\acknowledgments
We thank J.~A.~Urrea-Ni\~no for many insights around the implementation of distillation profiles, A.~Risch for the development of the perambulator code and members of the FOR5269 research unit for useful discussions. This work made use of QUDA \cite{Clark:2009wm} for performing inversions and qcdlib for the calculation of Laplacian eigenvectors. The authors gratefully acknowledge the computing time provided on the high-performance computer HoreKa by the National High-Performance Computing Center at KIT (NHR@KIT). This center is jointly supported by the Federal Ministry of Research, Technology and Space and the Ministry of Science, Research and the Arts of Baden-Württemberg, as part of the National High-Performance Computing (NHR) joint funding program. HoreKa is partly funded by the German Research Foundation (DFG). The authors also gratefully acknowledge the scientific support and HPC resources provided by the Erlangen National High Performance Computing Center (NHR@FAU) of the Friedrich-Alexander-Universität Erlangen-Nürnberg (FAU) under the NHR project k103bf. LB acknowledges support from the Agence Nationale de la Recherche (ANR) under Project No. ANR-23-CE31-0019 and the German Research Foundation (DFG) research unit FOR5269 ``Future methods for studying confined gluons in QCD". This project received funding from the European Research Council (ERC) via the project ``LEEX” grant agreement 101170304, funded by the European Union. Views and opinions expressed are however those of the author(s) only and do not necessarily reflect those of the European Union or the European Research Council Executive Agency (ERCEA). Neither the European Union nor the ERCEA can be held responsible for them. 

\bibliographystyle{JHEP}
\bibliography{refs}

\end{document}